\documentclass{iaus}
\usepackage{graphics}

  \checkfont{eurm10}
  \iffontfound
    \IfFileExists{upmath.sty}
      {\typeout{^^JFound AMS Euler Roman fonts on the system,
                   using the 'upmath' package.^^J}%
       \usepackage{upmath}}
      {\typeout{^^JFound AMS Euler Roman fonts on the system, but you
                   dont seem to have the}%
       \typeout{'upmath' package installed. iaus.cls can take advantage
                 of these fonts,^^Jif you use 'upmath' package.^^J}%
      }
  \else
  \fi

  \checkfont{msam10}
  \iffontfound
    \IfFileExists{amssymb.sty}
      {\typeout{^^JFound AMS Symbol fonts on the system, using the
                'amssymb' package.^^J}%
       \usepackage{amssymb}%

      }{}
  \fi

  \IfFileExists{amsbsy.sty}
    {\typeout{^^JFound the 'amsbsy' package on the system, using it.^^J}%
     \usepackage{amsbsy}}
    {}

\newsavebox{\astrutbox}
\sbox{\astrutbox}{\rule[-5pt]{0pt}{20pt}}

\newcommand\etal{\mbox{\textit{et al.}}}

\title[The Interplay among Black Holes, Stars and ISM in Galactic 
       Nuclei]{Feeding AGN: new results from the NUGA survey}
\author[S. Garc\'{\i}a-Burillo {\it et al.\/}]%
{S. Garc\'{\i}a-Burillo$^1$, F. Combes$^2$, E. Schinnerer$^3$, F. Boone$^4$, L.~K. Hunt$^5$, A. Eckart$^6$, L.~J. Tacconi$^7$, S. Leon$^8$, A.~J. Baker$^7$, P. Englmaier$^9$, \and R. Neri$^{10}$}
\affiliation{$^1$Observatorio Astron\'omico Nacional (OAN), Madrid, Spain, e-mail:s.gburillo@oan.es\\[\affilskip]
$^2$Observatoire de Paris, LERMA, Paris, France, e-mail:francoise.combes@obspm.fr\\[\affilskip]
$^3$NRAO, Socorro, USA, e-mail:eschinne@nrao.edu\\[\affilskip]
$^4$Bochum University, Bochum, Germany, e-mail:fboone@astro.rurh-uni-bochum.de\\[\affilskip]
$^5$Instituto di Radioastronomia/CNR,Firenze, Italy, e-mail:hunt@arcetri.astro.it\\[\affilskip]
$^6$Universit\"at zu K\"oln, K\"oln, Germany, e-mail:eckart@ph1.uni-koeln.de\\[\affilskip]
$^7$MPE, Garching, Germany, e-mail:linda@mpe.mpg.de, ajb@mpe.mpg.de\\[\affilskip]
$^8$Instituto de Astrof\'{\i}sica de Andaluc\'{\i}a, Granada, Spain, e-mail:stephane@iaa.es\\[\affilskip]
$^9$Universit\"at Basel, Binningen, Switzerland, e-mail:ppe@astro.unibas.ch\\[\affilskip]
$^{10}$IRAM, Grenoble, France, e-mail:neri@iram.fr}

\pubyear{2004}
\volume{222}
\pagerange{1--8}
\date{?? and in revised form ??}
\jname{The Interplay among Black Holes, Stars and ISM \\in Galactic Nuclei}
\editors{Th. Storchi Bergmann, L.C. Ho \& H.R. Schmitt, eds.}
\begin{document}

\maketitle

\begin{abstract}
The NUGA project is a high-resolution (0.5''-1'') CO survey of low luminosity AGN including the full sequence of activity types (Seyferts, LINERs and transition objects). NUGA aims to systematically study the different mechanisms for gas fueling of AGNs in the Local Universe. In this paper we discuss the latest results of this recently completed survey, which now includes newly acquired subarcsec resolution observations for all targets of the sample. The large variety of circumnuclear disk morphologies found in NUGA galaxies ($m=1$, $m=2$ and stochastic instabilities) is a challenging result that urges the refinement of current dynamical models. In this paper we report on new results obtained in 4 study 
cases for NUGA: NGC\,4826, NGC\,7217, NGC\,4579 and NGC\,6951.
\end{abstract}

\firstsection 
\section{The NUGA project: searching for AGN feeding mechanisms}

While it is widely accepted that the onset of activity results from the feeding of a supermassive black hole by the gas reservoir from its host galaxy, there is no consensus on which mechanisms can remove virtually  all of the gas angular momentum and drive the infall down to scales of tens of pc. Furthermore, it is unknown whether these mechanisms are at work only in active galaxies, or, alternatively, if the key
difference between AGN and {\it pure} starburst or {\it quiescent} galaxies resides in the availability of gas to supply the central engine. The NUclei of GAlaxies--NUGA--project (\cite[Garc\'{\i}a-Burillo \etal\ 2003a; Garc\'{\i}a-Burillo \etal\ 2003b]{cont1, paper1}) is the first high-resolution ($\sim$0.5$^{\prime\prime}$--1$^{\prime\prime}$) CO survey of 12 low luminosity AGN (LLAGN) including the full sequence of activity types (Seyferts, LINERs and transition objects). Observations were carried out with the IRAM Plateau de Bure Interferometer (PdBI). Thus far, no interferometric survey had been focused on the study of AGN in the Local Universe (e.g., \cite[Helfer \etal\ 2003]{hel03}). High-resolution observations are paramount to achieve a sharp view of the distribution and kinematics of molecular gas down to the critical scales for AGN feeding ($<$100~pc). On these scales, {\it secondary} modes, embedded in kpc-scale perturbations (e.g., bars, spirals and tidal instabilities), are expected to take over. Furthermore, to get a complete picture of the AGN feeding issue NUGA relies on a multi-wavelength approach. Information on the stellar potentials and the star formation processes in NUGA targets is obtained from available HST and ground-based optical/NIR images. This multi-wavelength strategy allows us to test the efficiency of AGN feeding mechanisms by a direct determination of the stellar gravitational torques exerted on the gas and, also, by numerical simulations of the gas dynamics.

The first images issued from NUGA reveal a wide variety of morphologies in the circumnuclear molecular 
disks of AGN hosts (\cite[Garc\'{\i}a-Burillo \etal\ 2003a; Garc\'{\i}a-Burillo \etal\ 2003c]{cont1,cont2}). Various gravitational instabilities develop at different spatial scales, including $m$=2, $m$=1 and stochastic perturbations. Some galaxies host several coexisting instabilities, while others host mainly one type of instability. The weak correlation between activity type and nuclear morphology of the AGN host suggests that the AGN duty cycle in LLAGNs is shorter than the relevant time scale of the instabilities present in the maps. The relevant question to be addressed is to what extent the identified perturbations favour AGN feeding or, eventually, if some of them can inhibit the process. As an illustration of the non-obvious quest for evidence of ongoing feeding in AGNs, we report below on recent results obtained in 4 study cases for the NUGA project: NGC\,4826, NGC\,7217, NGC\,4579 and NGC\,6951.

\begin{figure}
 \includegraphics{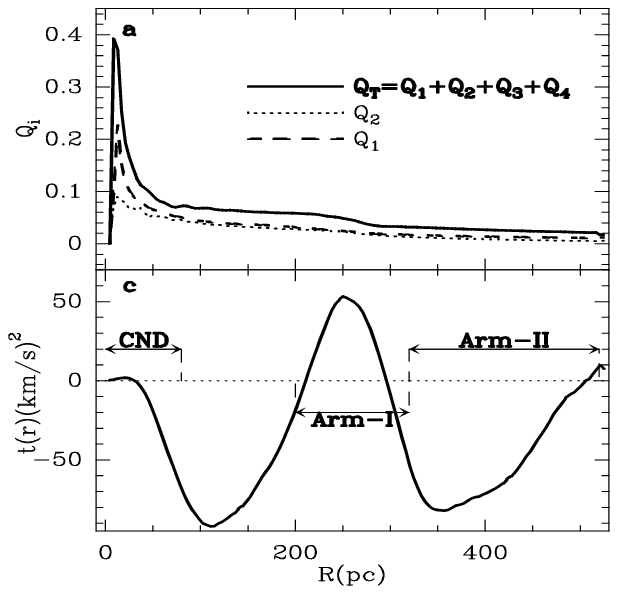}
 \includegraphics{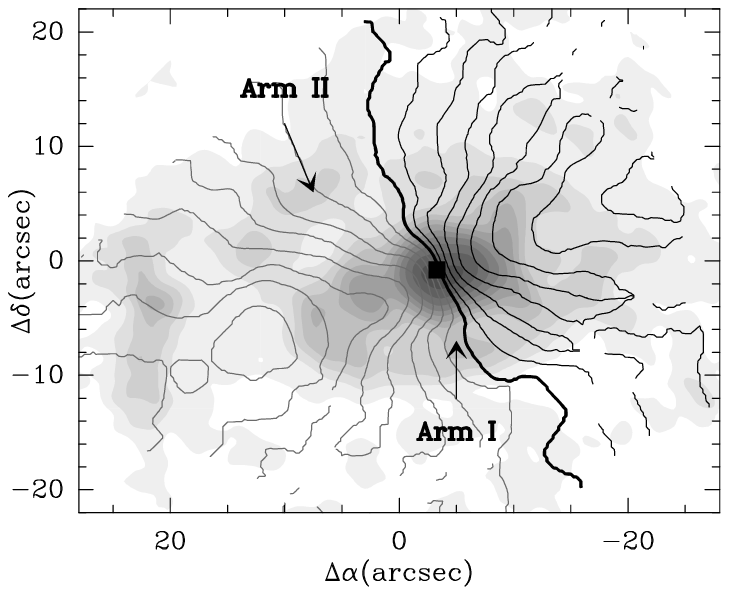}
 \vspace{-4.2cm}
 
 \caption{Gravitational stellar torques on molecular gas evaluated in the
600\,pc disk of the LINER NGC\,4826 (adapted from \cite[Garc\'{\i}a-Burillo \etal\ 2003b]{paper1}). The strength of the i-Fourier components of the stellar potential, Q$_i$, derived from HST-NICMOS data is shown in the {\bf upper left} panel. The  {\bf lower left} panel shows the radial variation of the average torque per unit mass (t(r)). The radial extent of Arm I and Arm II, both indicated in the {\bf right} panel (which overlays isovelocity contours onto the $^{12}$CO(1--0) intensity map of NGC\,4826), is also given. AGN feeding seems to be inefficient at present (see text).}
\label{fig:torques}

\end{figure}
\begin{figure}
 \includegraphics{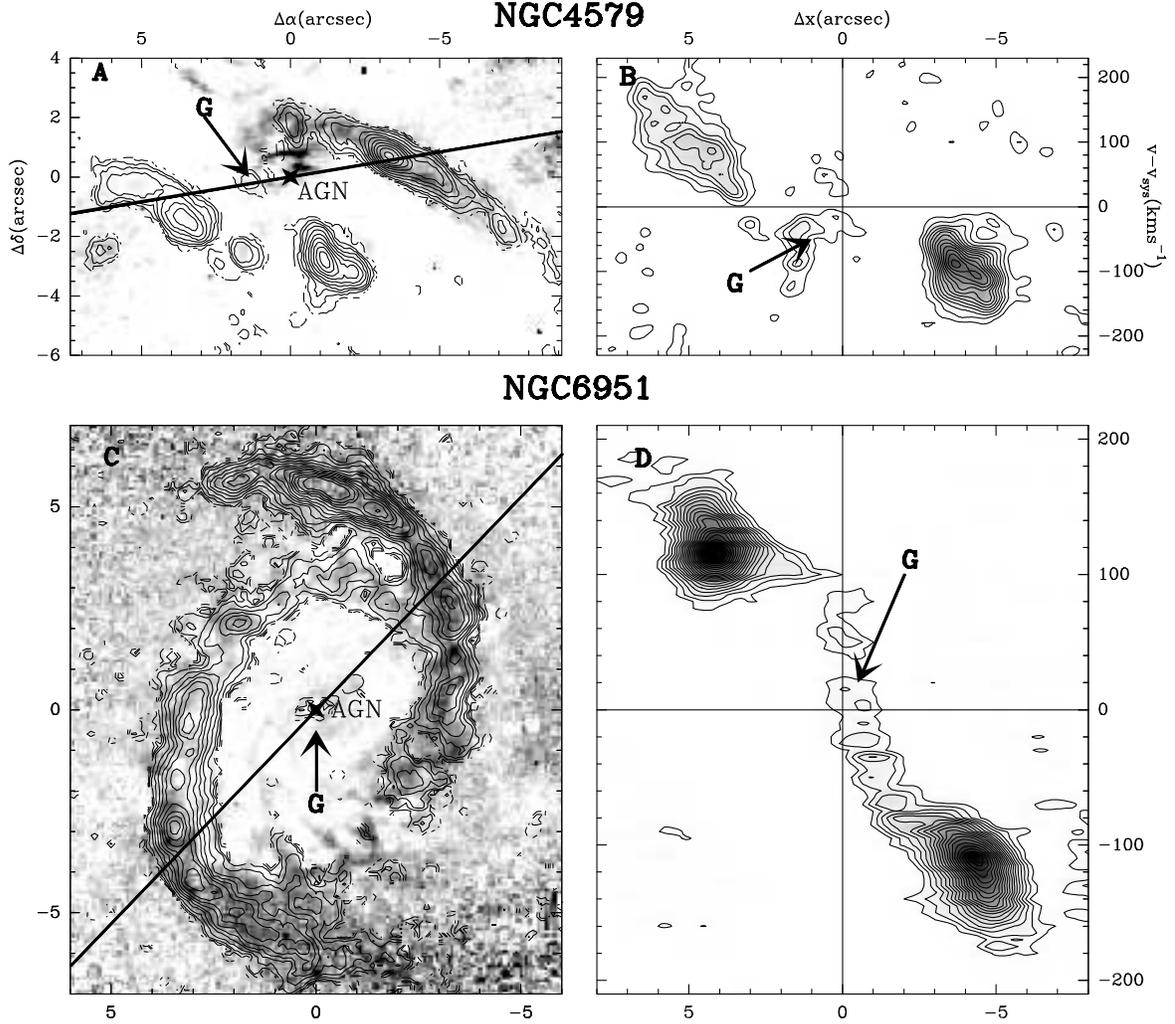}
 \caption{We overlay $^{12}$CO(2--1) contours on HST NIR color images (darker shades for redder colors) for two Seyferts in the NUGA sample observed at subarcsec resolution at the PdBI: NGC\,4579 (L1.9/Sy1.9) (Garc\'{\i}a-Burillo \etal 2004, in prep) and NGC\,6951 (Sy2) (Schinnerer \etal 2004, in prep), appearing in panels {\bf A} and {\bf  C}, respectively.  The locations of the AGN are given by star markers; the detections of $\sim$10$^{6}$M$_{\odot}$ molecular gas components near the AGN (denoted as G) are indicated by the arrows. The kinematics of the gas components {\it very near} the AGN are dissimilar in these two Seyferts: highly non-circular in NGC\,4579, and mostly circular in NGC\,6951.
  This is illustrated in panels {\bf B} and {\bf D} showing the major axis p-v plots (oriented as shown in left panels) for NGC\,4579 and NGC\,6951, respectively. In both cases, NUGA maps show tantalizing evidence of what could be ongoing AGN feeding episodes.}\label{fig:feeding}
\end{figure}


\section{AGN feeding in LINERs: NGC\,4826 and NGC\,7217}

The circumnuclear molecular gas disk of the LINER/transition object NGC\,4826 shows the prevalence of $m$=1 instabilities (see figure~\ref{fig:torques}, adapted from \cite[Garc\'{\i}a-Burillo \etal\ 2003b]{paper1}). Gas kinematics are perturbed by streaming motions related to $m$=1 {\it modes}. The non-circular motions associated with the inner $m$=1 perturbations (lopsided instability and inner $m=1$ spiral--Arm-I) fit the pattern expected for a trailing wave developed outside corotation ('fast' wave). A paradoxical
consequence is that the inner $m$=1 perturbations would not favour AGN feeding. An independent
confirmation that the AGN is not being generously fueled at present is found in the low values of
the gravitational torques exerted by the stellar potential for R$<$530~pc (figure~\ref{fig:torques}).
While the radial variation of the stellar torques seem to account qualitatively for the changing
signature of streaming motions, the maximum value of the mean torque is exceedingly small: $\sim$50$(km/s)^2$. The combination of the stellar perturbations seems to make the gas lose its angular momentum rather inefficiently (see \cite[Garc\'{\i}a-Burillo \etal\ 2003b]{paper1} for details).

NUGA observations of the LINER NGC\,7217 have been compared with numerical simulations in \cite{paper2}. The emerging scenario in this LINER is {\it apparently} in direct contrast with what is found for NGC\,4826: in NGC\,7217, molecular gas is piled up in an axisymmetric ring and kinematics is seen to be dominated by regular rotation. N-body simulations developed in \cite{paper2} served to monitor the formation and evolution of the gas ring. Results from these simulations show that gas inside the ring experiences positive gravity torques from the stellar oval perturbation which builds up the resonant ring near the ILR; hence, gas in the inner 700\,pc is experiencing an outward flow. As for NGC\,4826 (but for totally different reasons!) we find no evidence of significant ongoing fueling of the NGC\,7217 nucleus.
Only a GMC-like non self-gravitating unit coincident with the AGN seems to betray a past accretion episode.

\section{AGN feeding in Seyferts: NGC\,4579 and NGC\,6951}

The S1.9/L1.9 barred galaxy NGC\,4579 has been mapped at $\sim$0.5$^{\prime\prime}$ resolution (see figure~\ref{fig:feeding}\textit{a,b} adapted from Garc\'{\i}a-Burillo \etal\ 2004, in prep). The molecular gas distribution in the inner 500\,pc mimics the gas response to a bar potential. The 
$^{12}$CO(2--1) map shows 1.6$\times$10$^{8}$M$_{\odot}$ of molecular gas piled up in 2 leading lanes. However, the $m=2$ point-symmetry in the gas distribution breaks up in the inner 100\,pc disk where lopsidedness seems to take over. As illustrated by figure~\ref{fig:feeding}\textit{a}, the distribution of neutral gas close to the AGN consists of a 150\,pc off-centered ringed disk which is also identified in the V--I color HST image of the galaxy. There is little molecular gas near the AGN: the closest molecular complex (at a radius r$\sim$80\,pc) has a mass of $\sim$a few 10$^{5}$M$_{\odot}$. Modelling work (in progress) is
key to interpreting the complex kinematics of molecular gas near the AGN, characterized by the presence of highly non-circular motions (figure~\ref{fig:feeding}\textit{b}). Results from numerical simulated orbits will shed light on the still controversial feeding issue in this Seyfert.

NGC\,6951 is a prototypical Seyfert 2 galaxy for which subarcsec resolution CO maps have been recently completed within the NUGA project (figure~\ref{fig:feeding}\textit{c,d} adapted from Schinnerer \etal\ 2004, in prep). Molecular gas distribution in the inner 700\,pc shows two highly contrasted nuclear spiral arms containing a significant gas reservoir of 4$\times$10$^{8}$M$_{\odot}$ which is presently feeding an ongoing star forming episode, also identified by its prominent radio-continuum emission (\cite[Ho \& Ulvestad 2001]{ho01}). The $m=2$ gas instability is likely reflecting gas crowding along the x$_2$ orbits of the stellar bar. As can be guessed from figure~\ref{fig:feeding}\textit{c}, only a small amount of molecular gas seems to have succeeded in making its way down to the AGN. A compact molecular complex of $\sim$a few 10$^{6}$M$_{\odot}$ appears to be linked to the central engine. Furthermore, there is a northern molecular gas component related to the filamentary dusty spiral seen in the J--H color HST image. This stochastic instability might be efficient enough to feed the AGN if it has overcome the likely positive stellar gravity torques inside the ILR.


\begin{thebibliography}{}


\bibitem[Combes \etal\ (2004)]{paper2}
{Combes, F., Garc\'{\i}a-Burillo, S., Boone, F. \etal} 2004 \textit{A\&A}, \textbf{414}, 857--872

\bibitem[Garc\'{\i}a-Burillo \etal\ (2003a)]{cont1} 
{Garc\'{\i}a-Burillo, S., Combes, F., Eckart, A. \etal} 2003a \comment{The NUclei of GAlaxies (NUGA) survey.} In \textit{Active Galactic Nuclei: from Central Engine to Host Galaxy}, ed by S. Collin, F. Combes, and I. Shlosman (ASP Conference Series, San Francisco 2003) \textbf{Vol 290}, 423--426

\bibitem[Garc\'{\i}a-Burillo \etal\ (2003b)]{paper1}
{Garc\'{\i}a-Burillo, S., Combes, F., Hunt, L.~K. \etal} 2003b \textit{A\&A}, \textbf{407}, 485--502 

\bibitem[Garc\'{\i}a-Burillo \etal\ (2003c)]{cont2} 
{Garc\'{\i}a-Burillo, S., Combes, F., Eckart, A. \etal} 2003c \comment{NUGA: 
the IRAM survey of low-L AGNs.} In \textit{The 4th Cologne-Bonn-Zermatt Symposium: the Dense Interstellar Medium in Galaxies} , ed. by S. Pfalzner, C. Kramer, C. Straubmeier, and A. Heithausen
(Springer Verlag, 2004) in press, astro-ph/0310797

\bibitem[Helfer \etal\ (2003)]{hel03}
{Helfer, T.~T, Thornley, M.~D., Regan, M.~W. \etal} 2003 \textit {ApJS}, \textbf{145}, 259-327 

\bibitem[Ho \& Ulvestad (2001)]{ho01}
{Ho, L.~C., Ulvestad, J.~S.} 2001 \textit {ApJS}, \textbf{133}, 77-118 








  
\end{thebibliography}
\end{document}